# Experimental Demonstration of Nonlinear Photoconductive Gain in N-Doped β-Ga$_2$O$_3$ Devices

Vikash K. Jangir, Graduate *Student Member, IEEE* and Sudip K. Mazumder, *Fellow, IEEE*

*Abstract*— Photoconductive devices, based on ultra-wide-bandgap (UWBG) materials, offer a promising pathway toward compact, high-voltage (HV) optoelectronic and optical sensing in harsh environments. In this Letter, we report field-tunable nonlinear photoconductive gain in vertical **β-Ga$_2$O$_3$** photoconductive devices under sub-bandgap visible-light excitation. Devices fabricated on a 5.6-μm-thick nitrogen-doped semi-insulating **β-Ga$_2$O$_3$** epilayer grown on a conductive Sn-doped substrate were characterized under 445-nm continuous-wave illumination. A distinct transition from linear to nonlinear gain photoconductive behavior is observed at a threshold electric field of approximately 0.67 MV/cm, resulting in approximately 20 times enhancement in photocurrent. Complementary TCAD simulations indicate strong electric-field localization and a rapid increase in impact-ionization generation at high bias, suggesting that impact-ionization–assisted carrier multiplication contributes to the observed gain. These results demonstrate a high-field visible-light photoconductive detection mode in β-Ga$_2$O$_3$ enabled by defect-assisted transport, providing a pathway toward field-tunable gain photodetectors operating without deep-ultraviolet excitation.

## I. Introduction

β-Ga$_2$O$_3$ is an UWBG semiconductor with a large bandgap (~4.8 eV) and a high theoretical critical electric field, making it attractive, compared to conventional Si- and GaAs-based devices, for high-field optoelectronics and photoconductive detection in harsh environments [1][2]. Most β-Ga$_2$O$_3$ photodetectors rely on band-to-band excitation in the deep-ultraviolet spectral range, which necessitates light sources that are difficult to integrate due to their size, optical delivery requirements, and packaging constraints [3][4]. In contrast, sub-bandgap visible-light detection can be enabled through optical ionization of deep defect states, allowing photoconductive operation under visible-wavelength laser excitation.

However, most reported β-Ga$_2$O$_3$ photoconductive devices operate in the linear photoconductive regime, where the photocurrent scales linearly with absorbed optical power [5]. This one-to-one photon–electron relationship necessitates high-energy optical excitation to achieve appreciable photocurrent levels, thereby limiting power gain and system efficiency. To overcome this limitation, photoconductive devices must operate in a nonlinear high-gain regime, where field-assisted carrier multiplication mechanisms enable a low-energy optical trigger to control substantially higher current levels. While such regimes are well established in GaAs-based PCSS, they remain largely unexplored experimentally in β-Ga$_2$O$_3$ [6][7].

In this letter, we present the first experimental demonstration of nonlinear high-field photoconductive gain in β-Ga$_2$O$_3$ devices. *N*-doped vertical β-Ga$_2$O$_3$ devices are optically excited using a 445-*n*m continuous-wave (CW) laser, revealing a distinct transition from linear photoconductive regime to a nonlinear high-field transport regime involving trap-assisted carrier release and impact-ionization effects at a threshold electric field of approximately 0.67 MV/cm. By leveraging deep-level trap states activated under sub-bandgap illumination, a current multiplication factor of ~20 and peak photocurrents of up to 475 μA are achieved.

## II. Device Structure and Experimental Set-up

Fig. 1 shows the schematic cross-section of the fabricated devices. The devices were fabricated on a 5.6 μm thickness *N*-doped (~2 × 10$^{16}$ cm$^{-3}$) semi-insulating epitaxial layer grown by Kyma technology using halide vapor phase epitaxy (HVPE) in a custom reactor designed and built by Kyma Technologies, Inc., following growth details as laid out in [8]. The substrate was a highly conductive Sn-doped 2" (001) β-Ga$_2$O$_3$ wafer procured from Novel Crystal Technologies, Inc. The growth was conducted using O$_2$ as the group VI precursor, with metallic Ga and Cl$_2$ forming GaCl *in-situ* as the group III precursor and utilizing NCl$_3$ for N-doping. The fabricated device size is 650 μm x 750 μm.

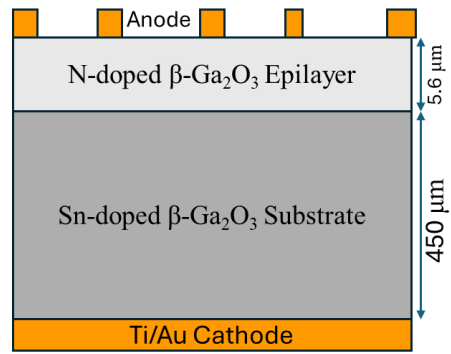

Fig. 1. Schematic cross-section of *N*-doped β-Ga$_2$O$_3$ device.

Ti/Au contacts (20 *n*m/100 *n*m) were used on both sides: a uniform bottom electrode covering the entire backside, and an interdigitated top electrode with alternating metal fingers and

The information, data, or work presented herein was funded in part by the Advanced Research Project Agency-Energy (ARPA-E), U.S. Department of Energy, under Award Number DE-AR0001879. The views and opinions of authors expressed herein do not necessarily state or reflect those of the United States Government or any agency thereof.

The authors are with the Department of Electrical and Computer Engineering, University of Illinois Chicago, Chicago, IL 60607 USA (e-mail: vjangi3@uic.edu; mazumder@uic.edu).

transparent gaps to ensure uniform field distribution and optical access (as illustrated in Fig. 1). After deposition, the contacts were annealed at 480 °C for 1 min in $N_2$ to improve contact quality. A standard AZ 1518 photoresist lift-off process was used to define the electrode patterns.

Electrical measurements were performed on a probe station using a Keithley 2657A high-voltage source-measure unit with an 8020 high-power interface for safe low-current operation. Optical excitation was provided by a 3-W fiber-coupled 445-nm CW diode laser. The optical fiber has a core diameter of 105 $\mu$m and a numerical aperture (NA) of 0.22. To prevent premature air breakdown and suppress surface flashover at high fields, the devices were immersed in a dielectric fluid (Fluorinert) during testing. All measurements, using the experimental setup shown in Fig. 2, were conducted at room temperature.

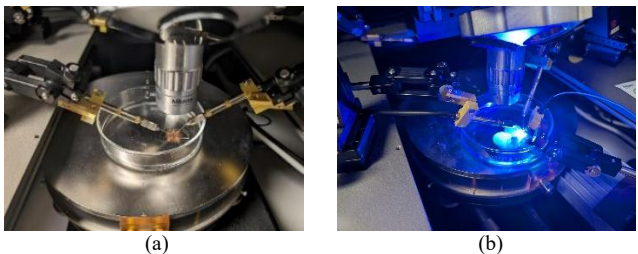

Fig. 2. Photocurrent measurement test set-up (a) off- and (b) on-states.

### III. RESULTS AND DISCUSSION

Fig. 3 shows the dark current and photocurrent characteristics of two vertical β-Ga₂O₃ devices. In Fig. 3(a), both devices exhibit a low dark current in the sub-μA range up to ~300 V, followed by a gradual increase at higher bias voltages. Under optical excitation, the photocurrent initially scales linearly with bias, corresponding to the linear photoconductive regime.

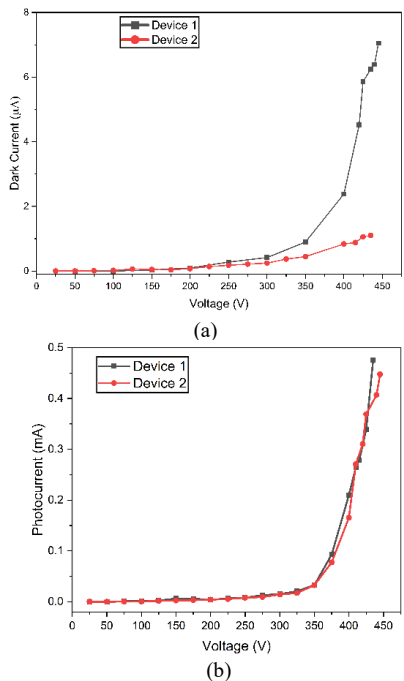

Fig. 3. Experimental I-V characteristics of the β-Ga₂O₃ devices. (a) Dark current showing the onset of trap ionization. (b) Photocurrent under 3-W, 445-*n*m illumination, illustrating the distinct transition from linear transport to the non-linear high-gain transport regime at 0.67 MV/cm.

Above approximately 325–350 V, a pronounced nonlinear increase in photocurrent is observed, reaching peak values of approximately 475 $\mu$A at 435 V in device 1 and approximately 450 $\mu$A at 445 V in device 2. Using 325 V as the linear-regime reference, the total photocurrent increases from approximately 20 $\mu$A to 476 $\mu$A at 435 V for device 1 and 450 $\mu$A at 445 V in device 2, corresponding to enhancement factors of approximately 23 and 21 times, respectively. The nonlinear increase in photocurrent with voltage, which is significantly larger than the corresponding increase in dark current, demonstrates a high photoconductive gain regime at elevated electric fields.

Furthermore, Figs. 4(a) and 4(b) show the temporal photocurrent response of the two devices under CW illumination at bias voltages of 300, 400, and 425 V. At low bias (300 V), the photocurrent exhibits a small and slowly varying response, consistent with limited carrier extraction under a moderate electric field. As the bias increases to 400 V and 425 V, a pronounced transient peak is followed by a gradual decay to a steady-state level. Both the peak amplitude and the steady-state photocurrent increase strongly with bias, indicating enhanced carrier generation and transport at high electric fields. The presence of a bias-dependent decay component suggests trap-mediated carrier dynamics, where rapid initial photo response is followed by carrier trapping and re-equilibration. The similar temporal behavior observed in both devices further confirms that the high-gain response arises from bulk transport mechanisms rather than measurement artifacts or contact-limited effects.

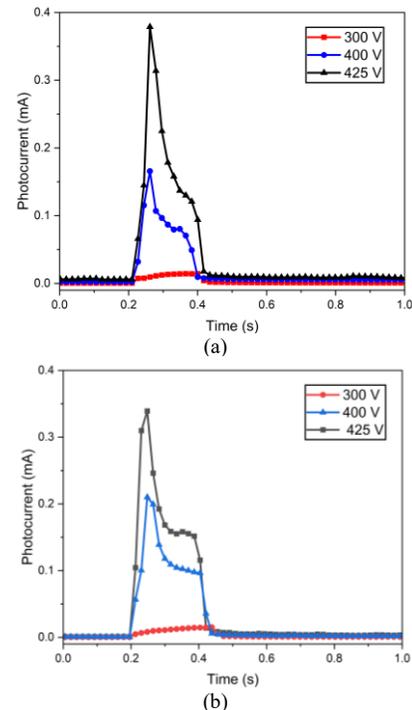

Fig. 4. Experimentally measured temporal photocurrent response of the β-Ga₂O₃ (a) device 1 and (b) device 2 measured on a probe station under CW laser illumination at bias voltages of 300 V, 400 V, and 425 V.

TCAD simulations were performed using Silvaco ATLAS to analyze the electric-field distribution and impact-ionization generation rate in the vertical β-Ga₂O₃ devices. Two-dimensional drift–diffusion simulations were carried out at 300 K. The device structure consisted of a 5.6-$\mu$m-thick β-Ga₂O₃ epilayer ($x$ = 0–650 $\mu$m, $y$ = 0–5.6 $\mu$m) grown on a conductive Sn-doped β-Ga₂O₃ substrate ($y$ = 5.6–10 $\mu$m). The epilayer was uniformly doped n-type with a background donor concentration of $N_D = 1.6 \times 10^{15}$ cm⁻³, representing residual Si doping, while the substrate was modelled with a high donor concentration of $N_D \approx 5 \times 10^{19}$ cm⁻³ to ensure low-resistance current spreading. The deep nitrogen acceptor trap with a density of $N_t = 1.9 \times 10^{16}$ cm⁻³ located at $E_C = -2.9$ $e$V and capture cross-sections $\sigma_n = \sigma_p = 2 \times 10^{-16}$ cm² [9] was included in the epilayer to model the semi-insulating behaviour. Shockley–Read–Hall recombination with Fermi–Dirac statistics, field-dependent mobility, and Auger recombination were enabled. Optical excitation was modelled using a normally incident 445-nm beam. High-field carrier multiplication was modelled using the Selber Herr impact-ionization model with parameters $A_n = A_p = 2.5 \times 10^6$ cm⁻¹ and $B_n = B_p = 3.96 \times 10^7$ V·cm⁻¹ [10]. All contacts were assumed ohmic.

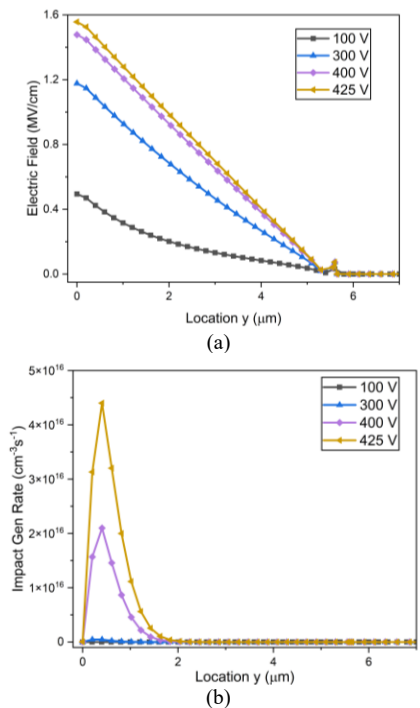

Fig. 5. (a) Simulated electric-field distribution along the vertical direction of the β-Ga₂O₃ device at applied biases of 100, 300, 400, and 425 V. (b) Corresponding impact-ionization generation rate profiles, showing a pronounced increase in generation at 400 V and above.

As shown in Fig. 5(a), the electric field increases monotonically with the applied bias and reaches peak values exceeding 1.5 MV/cm near the contact region at 400 V and 425 V, while gradually decreasing across the bulk of the substrate. Correspondingly, the simulated impact-ionization generation rate, as shown in Fig. 5(b), increases by several orders of magnitude at these biases, reaching peak values on the order of $10^{16}$–$10^{18}$ cm⁻³·s⁻¹. Although the impact generation is spatially localized near the high-field region, it provides an additional carrier multiplication pathway that enhances the overall photocurrent at high bias. These results indicate that avalanche multiplication contributes to the observed high-gain regime under CW illumination, acting in conjunction with bulk photoconductive transport mechanisms.

## IV. Conclusion

This letter presented the first experimental demonstration of nonlinear photoconductive gain operation in β-Ga₂O₃ devices under CW visible-light excitation. By employing a thin nitrogen-doped insulating epilayer on a conductive substrate, high electric fields were achieved at reduced operating voltages, enabling a transition from linear photoconduction to a high-gain regime at ~0.67 MV/cm. The resulting photocurrent exhibits an approximately 20-time enhancement relative to the linear regime, with peak values approaching 475 $\mu$A. Experimental transient measurements and TCAD simulations together suggest that this gain arises from the combined effects of trap-assisted transport and localized impact ionization. These results demonstrate the feasibility of visible-light-triggered β-Ga₂O₃ device for compact, high-field photoconductive detection applications without the need for deep-ultraviolet optical sources.